\documentclass[aip,cha,reprint,author-year]{revtex4-1}
\usepackage[english]{babel}
\usepackage{graphicx}
\usepackage{dcolumn}
\usepackage{bm}

\begin{document}

\title{Atmospheric rivers as Lagrangian coherent structures}

\author{Daniel Garaboa}\email{angeldaniel.garaboa@usc.es}
\author{Jorge Eiras-Barca}
\affiliation{Group of Nonlinear Physics, University of Santiago de Compostela, 15782 Santiago de Compostela, Spain.}
\author{Florian Huhn}
\affiliation{Institute of Mechanical Systems, ETH Z\"urich, Tannenstrasse 3, 8092 Z\"urich, Switzerland.}
\author{Vicente P\'erez-Mu\~nuzuri}\email{vicente.perez@cesga.es}
\affiliation{Group of Nonlinear Physics, University of Santiago de Compostela, 15782 Santiago de Compostela, Spain.}

\date{\today}

\begin{abstract}
We show that filamentous Atmospheric Rivers (ARs) over the Northern Atlantic Ocean are closely linked to attracting Lagrangian Coherent Structures (LCSs) in the large scale wind field. LCSs represent lines of attraction in the evolving flow with a significant impact on all passive tracers. Using Finite-Time Lyapunov Exponents (FTLE), we extract LCSs from a two-dimensional flow derived from water vapor flux of atmospheric reanalysis data and compare them to the three-dimensional LCS obtained from the wind flow. We correlate the typical filamentous water vapor patterns of ARs with LCSs and find that LCSs bound the filaments on the back side. Passive advective transport of water vapor from tropical latitudes is potentially possible.
\end{abstract}


\maketitle

\begin{quotation}
Atmospheric Rivers (ARs) transport moisture from the Tropics towards northern latitudes and they are responsible for extreme precipitation and flood events as they approach coastal areas. Advection of Lagrangian tracers by geophysical flows has shown the presence of persisting transport patterns of Lagrangian Coherent Structures (LCS).
In this paper, we argue that under some circumstances, ARs can be considered attracting LCS in front of which moisture accumulates, while the rivers propagate northwards.
This behavior has been observed both in three dimensional simulations of the wind flow and in terms of the two-dimensional vertically averaged water vapor field.
\end{quotation}

\section{Introduction}
The transport of moisture from the tropics to mid-latitudes is not continuous and uniform, but rather intermittent. More than 90\% of poleward water vapor is transported by narrow and elongated structures (longer than 2000\,km and narrower than 1000\,km), mostly within the Warm Conveyor Belt (WCB) ahead of cold fronts and within the Low Level Jet (LLJ) of extratropical cyclones commonly associated to the polar front \citep{ARs4,ARs6,Ralph2011}.
These structures, referred to as Tropospheric or Atmospheric Rivers (ARs), are defined as elongated regions of Integrated Water Vapor column (IWV) over $2$\,cm and winds stronger than $12\,\mathrm{m/s}$, that transport moisture in the lower troposphere close to the $850$\,hPa level \citep{Ralph2011,Neiman2008a,Leung2009}.
The advection of moisture by ARs is a key process for the Earth's sensible and latent heat redistribution and has a strong impact on the water cycle of the mid-latitudes by generating extreme precipitation events.
The connection between extreme precipitation and flood events has been shown over the Western US Coast \citep{Leung2009,Stohl2008} and over Europe \citep{Lavers2013}.

In large scale geophysical flows, advection is often the primary process shaping tracer patterns, and effects such as turbulent diffusion or other sinks and sources of the tracer are secondary.
Such tracer patterns have been observed in the atmosphere and in the ocean; examples include volcanic ash clouds \citep{Gudmundsson2012}, plankton blooms \citep{Toner2003,Lehahn2007,Huhn2012}, and oil spills \citep{Olascoaga2012}.
If advection is dominant, Lagrangian Coherent Structures (LCS) are the relevant finite-time structures in the time-dependent flow that determine the deformation of the fluid and hence the evolution of any advective tracer field. The concept of LCSs has its origin in dynamical systems theory. It has been introduced by \citet{Haller2000} and is still being developed further \citep{Farazmand2014}. Attracting hyperbolic LCSs are lines evolving with the flow that maximally attract fluid. In the vicinity of attracting LCSs, the fluid is stretched in one direction and compressed in the orthogonal direction. Therefore attracting LCSs are the cores of filamentous tracer patterns.
A standard way of estimating the position of attracting LCSs is the Finite-time Lyapunov Exponent (FTLE) \citep{Haller2002,Shadden2005}.
The FTLE measures the maximum stretching rate among trajectories over a fixed time interval derived from neighbor fluid particles. Ridges of maximum FTLE values are a robust method to estimate hyperbolic LCSs \citep{Haller2002,Shadden2005}, although some drawbacks have been reported \citep{Haller2011}.

FTLE and similar versions of the Lyapunov exponent have been used to estimate LCSs in different models of atmospheric flows, such as a zonal stratospheric jet \citep{BeronVera2010}, a jet-stream \citep{Tang2010}, a hurricane \citep{Rutherford2010}, transient baroclinic eddies \citep{Hardenberg2000} and the polar vortex \citep{Koh2002}.

Given that ARs over the Atlantic and Pacific Ocean appear as coherent filaments of water vapor with a persistence time of several days up to a week, and given that LCSs have turned out to explain the formation of similar tracer patterns in geophysical flows, this paper addresses the questions: Are ARs associated to attracting LCSs in the large scale tropospheric flow? To what extent do ARs consist of humid air advected from tropic latitudes?

Since ARs are defined and studied in terms of two-dimensional maps of IWV, we use a two-dimensional height-averaged tropospheric flow to detect the main LCSs. We focus on selected ARs events that propagate over the Northern Atlantic Ocean and hit the Iberian Peninsula. Wind fields and water vapor flux fields are retrieved from European Center for Medium-Range Weather Forecast reanalysis, ERA-Interim.
We focus on spatiotemporal patterns in the two-dimensional water vapor field formed during an AR event. These patterns are compared with advective structures in the wind field, namely LCSs.
As mentioned earlier, ARs are three-dimensional phenomena \citep{Sodeman2013}, so we have also considered a three-dimensional case study to address the limitations of a two-dimensional approach \citep{Stohl1998}.

We find that ARs with a sharp filamentous shape are associated to a strong attracting LCS in the wind field.
This suggests a dominance of horizontal advection in the formation process of the filamentous pattern. In these cases, the LCS typically marks the northern boundary of the AR - a line of high water vapor gradient where humid and dry air converge. Filamentous LCSs appear mainly in winter, while wide, less coherent ARs structures without a clear attracting LCS develop in summer.

\section{Methods}
Atmospheric Rivers have been analyzed in terms of a two-dimensional vertically integrated flow based on data retrieved from the European Center for Medium-Range Weather Forecast reanalysis, ERA-Interim \citep{Dee2011},
\begin{eqnarray}
Q&=&\frac{1}{g}\int\limits_0^1 q \frac{\partial p}{\partial \eta}d\eta  \label{eq:Q} \\
\Phi_\lambda&=&\frac{1}{g}\int\limits_0^1 uq \frac{\partial p}{\partial \eta}d\eta  \\ \label{eq:eastward}
\Phi_\phi&=&\frac{1}{g}\int\limits_0^1 vq \frac{\partial p}{\partial \eta}d\eta  \label{eq:northward}
\end{eqnarray}
where $Q$ is the vertically integrated water vapor, $\mathbf{\Phi}=(\Phi_\lambda,\Phi_\phi)$ are the eastern/northern water vapor flux, $g$ is the acceleration of gravity, and $\eta$ is a hybrid vertical coordinate \citep{ARs10,ARs9}. This coordinate uses the mean sea level as a bottom reference level and $p$ is the pressure level in the $\eta$ coordinate. $u$ and $v$ are the eastward and northward wind component and $q$ is the specific humidity. Data are available with a spatial resolution of $0.7^\circ$ and temporal resolution of 6 hours. The eastward and northward averaged velocities in a longitude-latitude spherical coordinate system ($\varphi,\theta$) are calculated as
\begin{eqnarray}
\mathbf{V_f}&=&\left[<\dot{\lambda}(\varphi,\theta,t)>,<\dot{\phi}(\varphi,\theta,t)>\right] \nonumber\\
&=&\left[\frac{\Phi_\lambda}{Q},\frac{\Phi_\phi}{Q}\right].
\label{eq:vflux}
\end{eqnarray}
When the velocity fields are vertically averaged in this way, the pressure levels are weighted according to their water vapor content, such that $\mathbf{V_f}$ is representative for the dynamics of ARs.

Along this paper, we will compare the results obtained with the previous vector field with those obtained from the wind velocity fields at individual pressure levels,
\begin{equation}
\mathbf{V_{w}}=\left[u(\varphi,\theta,t),v(\varphi,\theta,t),w(\varphi,\theta,t)\right]_p
\label{eq:vwind}
\end{equation}
Hence, two types of Lagrangian simulations have been performed. Either particle trajectories are computed at constant pressure levels (2D simulation), or a full 3D simulation is done. For the last case, the top boundary of the domain is located at 300~hPa, while the bottom boundary is at 1000~hPa. Particle trajectories are computed by integrating (\ref{eq:vflux}) or (\ref{eq:vwind}) using a 4-th order Runge-Kutta scheme with a fixed time step of $\Delta t=2$~hours, and a multilinear interpolation in time and space. A fine grid of particles with an initial separation of $1/5^\circ$ in each pressure level is advected to obtain fields of the Lagrangian quantities introduced below. For the two-dimensional simulations $w=0$ in Eq.~(\ref{eq:vwind}). The finite integration time is chosen to be $\tau = 120$ hours, a typical time scale for the formation and propagation of the ARs.

In order to detect attracting Lagrangian coherent structures, we use the Finite-Time Lyapunov Exponent (FTLE) field $\sigma(\mathbf{x})$ which is a measure of stretching about fluid trajectories. It is computed from the trajectories of Lagrangian tracers in the flow \citep{Peacock2010} as
\begin{equation}
\sigma(\mathbf{r_0},t_0,\tau) = \frac{1}{|\tau|}\,\log\sqrt{\mu_{max}(\mathbf{C}(\mathbf{r_0}))},
 \label{eq:lyapunov}
\end{equation}
where $\mu_{max}$ is the maximum eigenvalue of the right Cauchy-Green deformation tensor $\mathbf{C} = \mathbf{F}^{T}\mathbf{F}$. $\mathbf{F}$ is defined by $\mathbf{F}(\mathbf{r}_0)=\mathbf{\nabla}(\mathbf{r}(t_0+\tau))$, and $\mathbf{r}(t_0+\tau)$ is the final position of the tracers. The time variation of the FTLE field has been computed by following the same steps explained previously, but varying the initial time $t_0$ in fixed steps $\Delta t_0 = 6$ hours in order to release a new initial tracer grid. FTLE are computed in forward and backward time directions.

Finite-Time Lyapunov Exponents have been used extensively to quantify mixing and especially to extract persisting transport patterns in the flow, referred to as Lagrangian Coherent Structures (LCS). The FTLE at a given location measures the maximum stretching rate of an infinitesimal fluid parcel over the interval $[t_0,t_0+\tau]$ starting at the point $\mathbf{r_0}$ at time $t_{0}$. Ridges of the FTLE field are used to estimate finite time invariant manifolds in the flow that separate dynamically different regions. Repelling (attracting) LCSs for $\tau>0$ ($\tau<0$) can be thought of as finite-time generalizations of the stable (unstable) manifolds of the system. These structures govern the stretching and folding mechanism that control flow mixing. LCSs as FTLE ridges are extracted using the Hessian matrix and the gradient of $\sigma(\mathbf{r_0},t_0,\tau)$ \citep{Shadden2005,Sadlo2007}. To discard weak structures, a threshold criterion associated to the 90th percentile is used for the FTLE values and for 
the eigenvalues of the Hessian matrix, a fixed threshold is used for the minimal connected ridge area. For the 3D simulations, the LCSs are estimated using a high isosurface of the FTLE field.

\section{Results}
\subsection{2D results}
\begin{figure*}
\includegraphics[width=\linewidth]{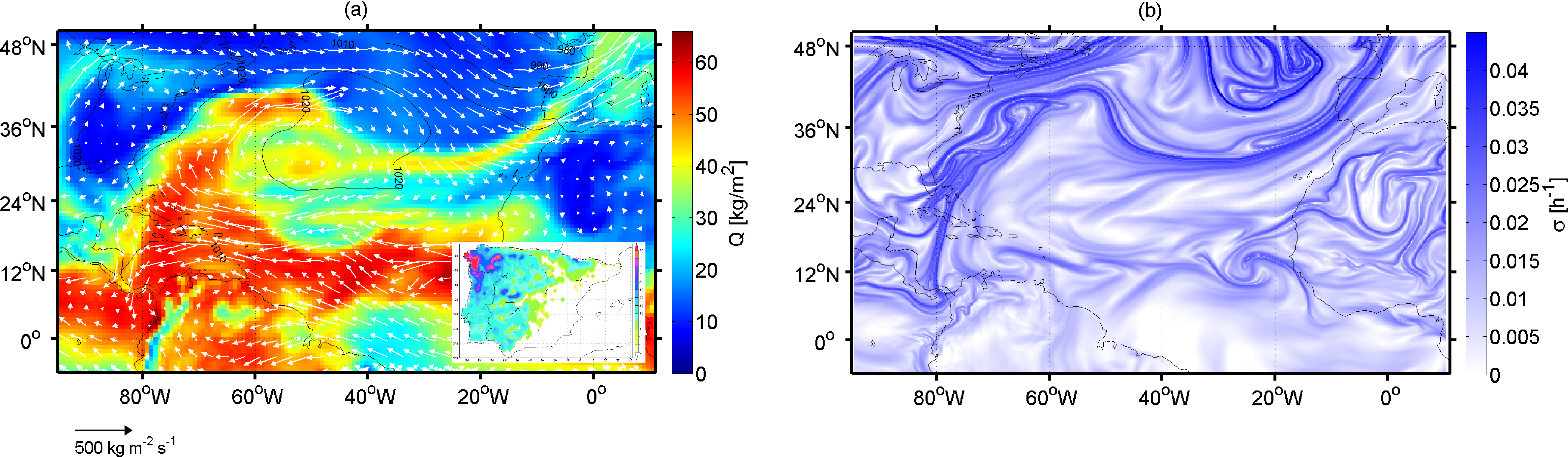}
\caption{(a) Atmospheric River event on 15 October 1987 in terms of the integrated water vapor column, Eq.~(\ref{eq:Q}). The $\mathbf{\Phi}$ flow is shown superimposed. The inset shows the accumulated precipitation rates over the Iberian Peninsula [mm]. (b) Backward FTLE field for the same date and the flow given by Eq.~(\ref{eq:vflux})}.
\label{fig:AR}
\end{figure*}
Atmospheric Rivers are observed as filamentous structures in fields of the integrated water vapor column $Q$, Eq.~(\ref{eq:Q}). Figure~\ref{fig:AR}(a) shows an example of an AR extending north-westwards from $\sim(20^\circ \mathrm{N}, 75^{\circ}\mathrm{W})$ and transporting water vapor over the Atlantic Ocean towards the Iberian Peninsula. The inset shows the daily precipitation rates well above the mean values for this area that are caused by this event. The long ridge of large values of $Q$ connects the tropics with the North Atlantic Ocean and represents a continuous AR structure. The backward FTLE field is shown in Fig.~\ref{fig:AR}(b). The FTLE ridge reaching from the Iberian Peninsula to the Gulf of Mexico shows a clear relationship to the Atmospheric River depicted previously. From a Lagrangian point of view, the AR can be considered as an attracting LCS or unstable manifold of the flow dynamical system $\mathbf{V_f}$ (\ref{eq:vflux}).

\begin{figure*}
\includegraphics[width=\linewidth]{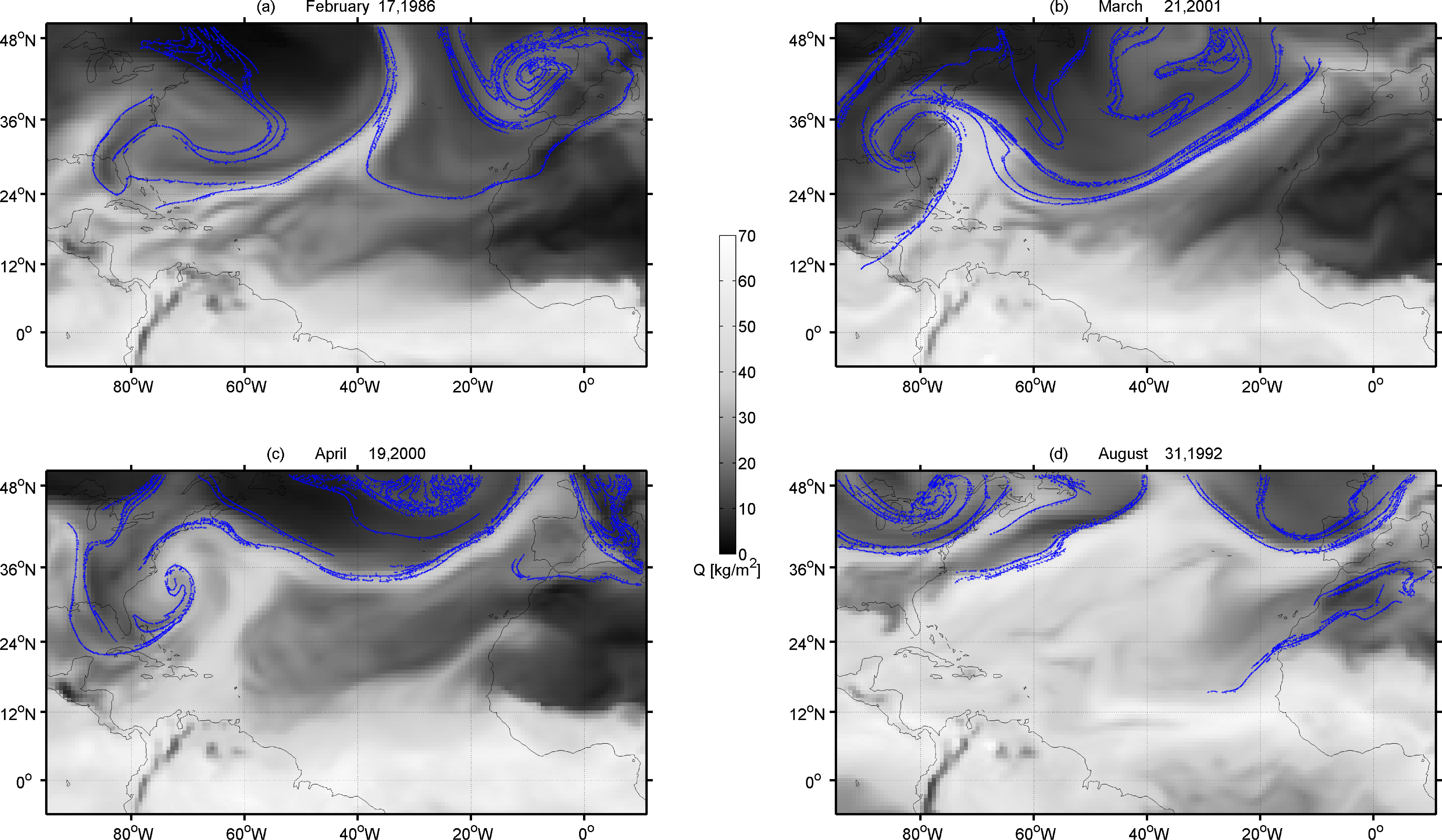}
\caption{Lagrangian Coherent Structures (blue dots) and integrated vapor column $Q$ for four different events. Note the coincidence between LCSs and the maximal $Q$ values in the first three cases (winter and spring).}
\label{fig:4ARs}
\end{figure*}

Similar results have been obtained for different dates. Lagrangian Coherent Structures calculated from the backward FTLE fields are compared to the $Q$ field (\ref{eq:Q}) in Fig.~(\ref{fig:4ARs}) for four ARs events. Note that regions of maximum $Q$ values match LCSs locations. The good agreement of LCSs computed from the flow field $\mathbf{V_f}$ with patterns in the tracer field $Q$ indicates that $\mathbf{V_f}$ captures the main dynamic processes shaping the water vapor field.

In three cases, Figs.~(\ref{fig:4ARs})(a) to (\ref{fig:4ARs})(c), LCSs and $Q$ ridges show a filamentous structure, while for the event in Fig.~(\ref{fig:4ARs})(d), LCSs have a scattered shape. Filamentous ARs have one or two narrow jets growing from low latitudes transporting water vapor to medium latitudes. However, potential ARs situations with a scattered shape do not correspond to well defined water vapor jets. Large areas of water vapor surround the jet. The last situation is typical of summer cases, while the first one occurs mainly in winter or beginning of spring. For the winter ARs cases, the threshold criteria used to extract LCSs
from the FTLE fields is insensitive to variations. However, small variations in the thresholds for the summer cases lead to very different LCSs patterns, an indication that the LCSs patterns associated to ARs are more coherent in winter than in summer.

In all analyzed winter ARs events, LCSs are located \emph{behind} the river (in the direction of propagation) and do not coincide with it. Here the shape of the river is defined as the ridge of the $Q$ field. From a Lagrangian point of view, the attracting LCS accumulates water vapor in front of the pattern moving towards the east, and a certain gap in between both maxima, LCSs and $Q$ field, is observed.

\begin{figure}
\includegraphics[width=\linewidth]{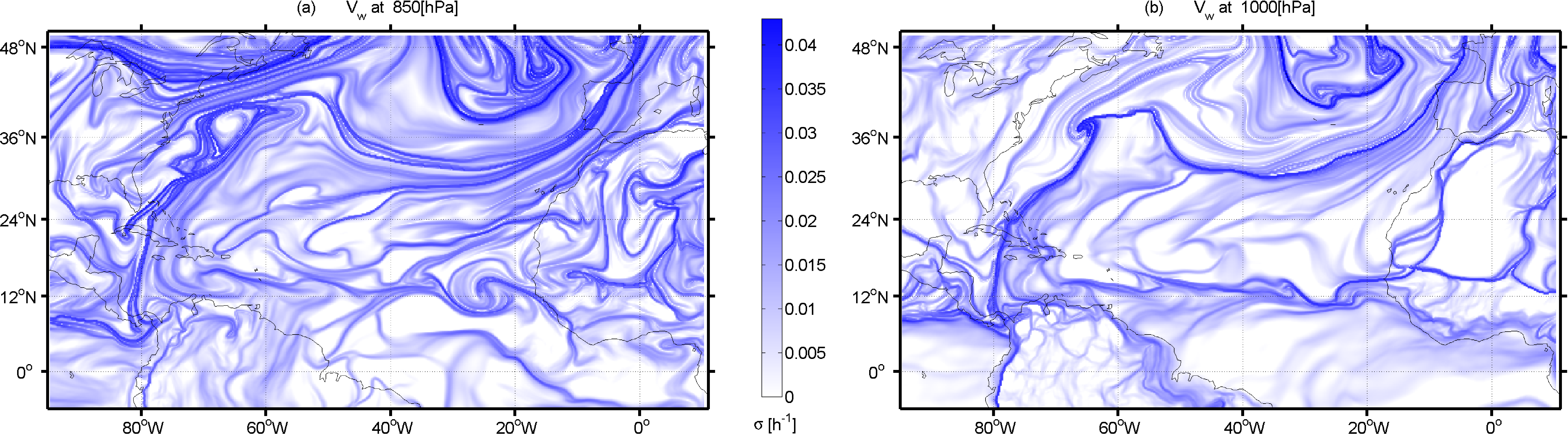}
\caption{FTLE backward fields computed using the wind field at two different pressure levels, 850 hPa (a) and 1000 hPa (b), for the ARs event on October 23, 1987.}
\label{fig:ARwind}
\end{figure}

These results are obtained using the vertically averaged flow $\mathbf{V_f}$ where the averaging is weighted with water vapor content. Another natural choice is to define a two-dimensional flow on isobaric levels.
We compute the FTLE fields at different individual isobars using the wind $\mathbf{V_w}$ (\ref{eq:vwind}) and neglecting the vertical wind component, $w=0$. Figure~\ref{fig:ARwind} shows an example of the FTLE fields obtained for the pressure levels 850\,hPa and 1000\,hPa. At 850 hPa, ridges of the FTLE field show a close relationship with those shown in Fig.~\ref{fig:AR}(b). For other pressure levels, we find that the main LCS preserve its continuity, but loses the smoothness of the curvature, showing a different shape.

\begin{figure}
\includegraphics[width=\linewidth]{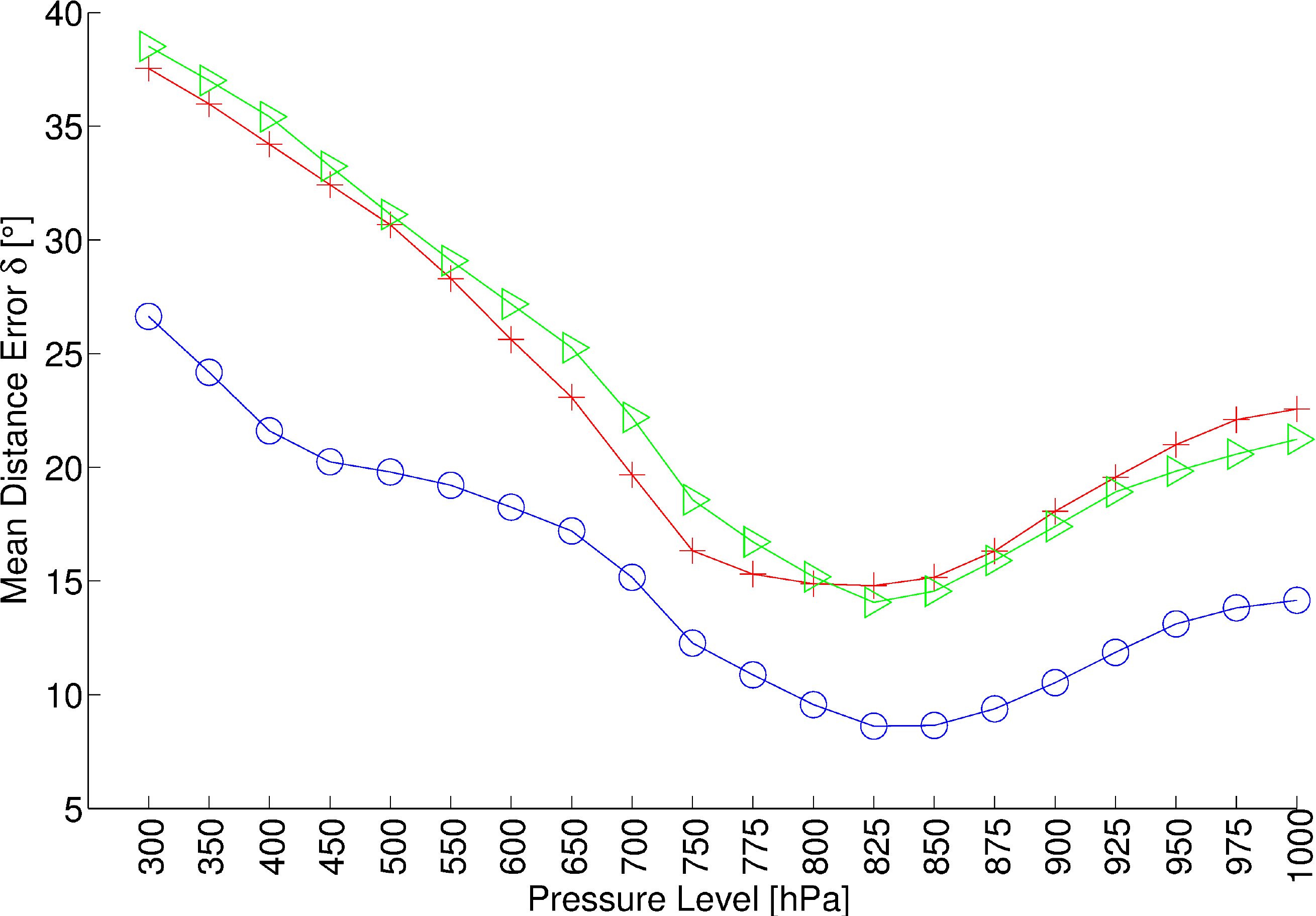}
\caption{Mean distance error defined by Eq.~(\ref{eq:relative}) for three ARs events at different pressure layers. December 1994 (red crosses), August 1988 (blue dots), and November 1995 (green triangles).}
\label{fig:error}
\end{figure}

However, the AR has a vertical extent and it is not clear which isobar to choose, such that the two-dimensional flow best represents the dynamics of the $Q$ tracer, i.e., such that is is most similar to $\mathbf{V_f}$.
In order to compare trajectories for both two-dimensional flow fields, $\mathbf{V_f}$ and $\mathbf{V_w}$, we introduce a similarity measure, the mean distance error $\delta$. It measures the mean arc length distance between the final position of the tracers obtained for both flows,
\begin{equation}
\delta = N^{-1} \sum_{i=1}^N|\mathbf{r_{v_w}} - \mathbf{r_{v_f}}|.
\label{eq:relative}
\end{equation}
The subindex of $\mathbf{r}$ accounts for the used flow field, and $N$ is the number of passive tracers released. The mean distance error $\delta$ is then calculated for 20 pressure levels from $300\,$hPa to $1000\,$hPa for different ARs events, as shown in Fig.~\ref{fig:error}. A minimum of $\delta$ is observed for a pressure level near 850 hPa for all ARs events analyzed.
The minimum suggests that the flow field at $850$\,hPa is closest to the water vapor flow $\mathbf{V_f}$.
It is consistent with observations that the core of ARs is typically located at that height \citep{Ralph2011,Neiman2008a,Leung2009}.


\subsection{3D results}
\begin{figure*}
\includegraphics[width=\linewidth]{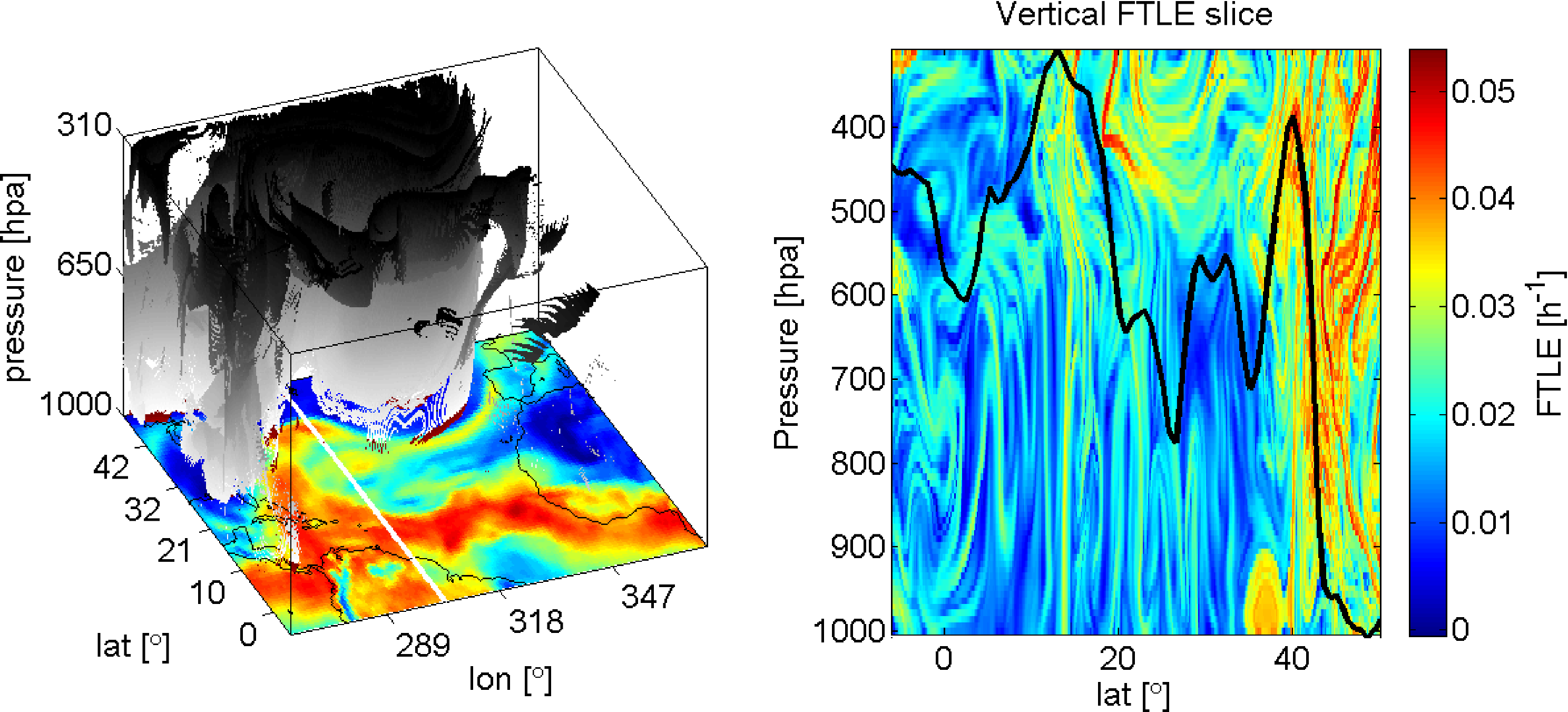}
\caption{Backward three-dimensional FTLE for the Atmospheric River event show in Fig.~(\ref{fig:AR}). The isosurface of the FTLE field of 0.04 h$^{-1}$ is shown in grayscale shades while the $Q$ field corresponding to this AR is depicted below. A vertical cut at $304^\circ$ longitude shows the presence of the AR at $40^\circ$ latitude. For comparison, $Q$ values at this longitude are shown superimposed over the FTLE field.}
\label{fig:LCS3D}
\end{figure*}

So far, LCSs have been calculated using the two-dimensional flow given by Eq.~(\ref{eq:vflux}).
However, including the vertical wind component might significantly alter the LCSs obtained from a purely two-dimensional analysis. To verify that, we compute the FTLE backward fields from a full 3D simulation using the wind field $\mathbf{V_w}$ (\ref{eq:vwind}). Figure~\ref{fig:LCS3D} shows the three-dimensional LCSs obtained using a fixed isosurface from the FTLE field for a winter AR. Note the ribbon extending from the Tropics towards the Iberian Peninsula. The surface of the LCS is continuous, independent of the pressure level, although slightly changes with increasing height. Note that the 3D FTLE isosurface follows the Atmospheric River depicted below in the figure. This fact suggests that the AR dynamics can be considered two-dimensional and the vector field defined by Eq.~(\ref{eq:vflux}) accounts for the global dynamics of the ARs. The vertical FTLE slice shows the presence of the atmospheric river at approximately $40^\circ$. For this latitude, $Q$ diminishes drastically, as the water vapor 
accumulates in front of the river as it displaces towards east. Note the accumulation of numerous filaments with large FTLE values \emph{behind} the AR due to flow mixing probably induced by the interaction between the jet stream and the AR \citep{Sodeman2013}.

\begin{figure}
\includegraphics[width=\linewidth]{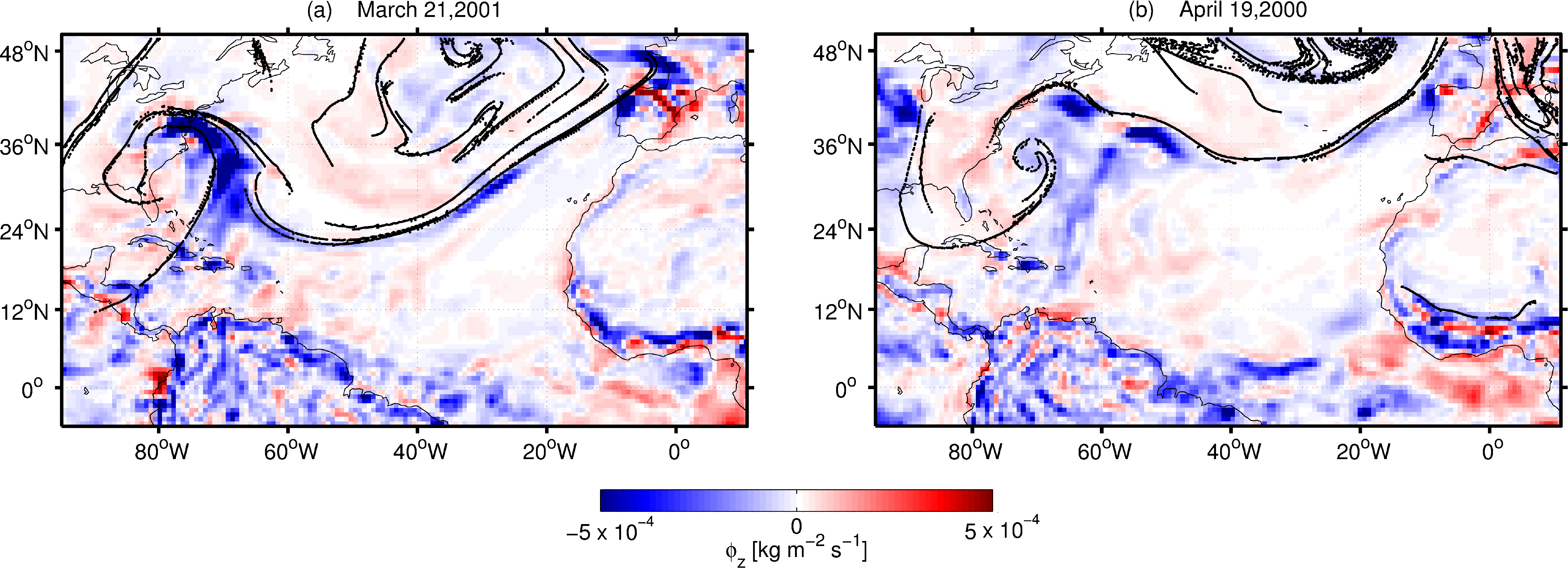}
\caption{Vertical flux of precipitable water at 850~hPa for two Atmospheric Rivers shown in Fig.~(\ref{fig:4ARs}). For comparison the LCSs detected in both cases are shown as black lines.}
\label{fig:FluxZ}
\end{figure}
Even if the horizontal shape of an AR turns out to be dominated by the horizontal flow, vertical motions in the atmosphere are needed to account for precipitation and evaporation processes. These processes may be represented by the vertical flux of precipitable water on a given level,
\begin{equation}
\phi_z = \rho\,w\,\left(q_v + q_l + q_s\right)
\label{eq:vertical}
\end{equation}
where $q_v$, $q_l$ and $q_s$ are the amount of water in the gaseous, liquid, and solid states, respectively, retrieved from the ERA-Interim database, $\rho$ is the density of the air and $w$ is the vertical component of wind. Figure~\ref{fig:FluxZ} shows the vertical flux of precipitable water for two days where ARs are present. The vertical flux (\ref{eq:vertical}) reaches maximum values all along the main attractive LCS barrier, but rather directly in front of the LCS than on top of it. This location of maximal vertical moisture flux corresponds to ridge of maximal water vapor content, cf. Fig.\,\ref{fig:FluxZ}\,b,c.
The 3D LCS seems to act as a lateral barrier for the vertical transformation processes of water vapor that develop in front of it as the AR moves eastward, showing that the LCS is a dynamical barrier to precipitable water.

One of the issues that emerges from these findings is that high vertical flux also means that $Q$ is not a conserved tracer, and if we want to characterize vertical motions of precipitable water active tracers are needed. But an implication of this is the possibility that the LCSs can identify regions which are not sensitive to the use of active tracers for horizontal movements of precipitable water.



\subsection{Origin of air}
\begin{figure}
\includegraphics[width=\linewidth]{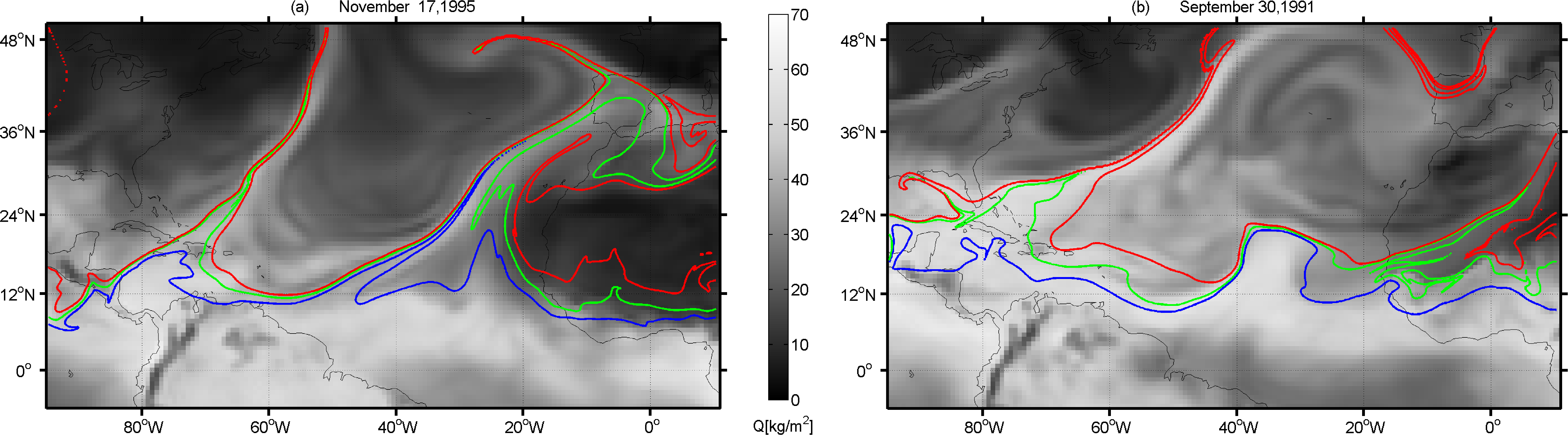}
\caption{Contours of latitudes of origin of passive tracers advected backward in time for two ARs events. The red, green and blue contours enclose particles coming from $25^{\circ}$N, $21^{\circ}$N, and $17^{\circ}$N. Shaded gray images correspond to water vapor concentration $Q$.}
\label{fig:ARorigin}
\end{figure}

Since we have observed stronger attracting LCSs for filamentous winter ARs than for scattered summer ARs, we also investigate the origin of air parcels in both cases. We quantify the displacement of passive particles for two ARs, once the AR hits the Iberian Peninsula. To that end, an initial grid of particles is advected backward in time for $\tau=5$\,days. Figure~\ref{fig:ARorigin} shows the origin of the particles in latitude for a filamentous (a) and a scattered (b) river. Note in panel (a) that particles coming from low latitudes coincide with high values of the integrated water vapor column.
This potentially gives the possibility that moisture from the Tropics is passively advected northwards, eventually forming the AR. The contribution of evaporation and precipitation can, however, not be assessed with this simple simulation of advection. Both processes tend to represent significant sinks and sources in the water vapor budget.
For the unstructured river (b), the contours of the original latitude do not enclose the ARs structure. Only a very thin zone from low latitudes with nearly zero area is coupled to the AR event. In this case, two-dimensional passive advection of moisture from low latitudes as a main source can be excluded.

\section{Conclusions}
The propagation of Atmospheric Rivers over the North Atlantic Ocean, finally hitting the Iberian Peninsula, has been studied in terms of Lagrangian tools. Based on an integrated water vapor flux obtained from the ERA-Interim database, AR events have been identified with attracting LCSs, both in 3D and 2D simulations. Two different AR events have been characterized.

On the one hand, narrow filamentous ARs with a fast and persistent eastward transport typically develop in winter. An attracting LCSs with the same shape as the ARs exists in the flow. It acts as a lateral boundary to the ARs. This boundary is more clearly visible for the 3D simulations where the AR is identified as a vertical ribbon displacing eastward.
As an advection experiment shows, for this kind of ARs, the air originates from low latitudes and passive transport of water vapor from the Tropics northwards is potentially possible.

On the other hand, for unstructured or scattered rivers, mostly occurring during the summer season, a clear attracting LCS seems to be absent. The water vapor balance of these ARs must be dominated by local sources, since the passive transport of moisture from low latitudes is practically excluded.

Our Lagrangian analysis assumes that water vapor behaves as a passive tracer. This assumption works to assess the deformation in the flow field and obtain LCSs that shape tracer patterns in the water vapor field, namely ARs. For a quantitative analysis of the water vapor balance in ARs, inertial tracers which may exchange matter and thermodynamic properties with the surrounding flow should be considered.

Finally, we conclude that the close connection between attracting LCSs and ARs should be taken into account for future studies and may help to characterize this kind of event. Most probably, our results are not limited to the Atlantic Ocean, but can be extended to ARs over the Pacific Ocean. An extensive analysis with more ARs events, including other regions, could help to set up a definition of ARs in terms of Lagrangian analysis.

\begin{acknowledgments}
ERA-Interim data was supported by the ECMWF. This work was financially supported by European Commission FP7 programme (EARTH2OBSERVE) through the project \emph{Global Earth Observation for Integrated Water Resource Assessment}, and by Ministerio de Econom\'{\i}a y Competitividad (CGL2013-45932-R). Computational part of this work was done in the Supercomputing Center of Galicia, CESGA. We acknowledge fruitful discussions with Drs. S.~Brands and G.~Miguez.
\end{acknowledgments}

\end{document}